\documentclass[12pt,preprint]{emulateapj}
\def\apj{{ApJ}}
\def\mnras{{ MNRAS}}

\def\be{\begin{equation}}
\def\ee{\end{equation}}
\def\bea{\begin{eqnarray}}
\def\eea{\end{eqnarray}}

\usepackage{graphicx}

\begin{document}

\title{Model-dependent estimate on the connection between fast radio bursts and ultra-high energy cosmic rays}

\author{Xiang Li\altaffilmark{1,2}, Bei Zhou\altaffilmark{1,2}, Hao-Ning He\altaffilmark{1}, Yi-Zhong Fan\altaffilmark{1}, and Da-Ming Wei\altaffilmark{1}}
\affil{$^1$ Key Laboratory of Dark Matter and Space Astronomy, Purple Mountain Observatory, Chinese Academy of Science,
Nanjing, 210008, China.}
\affil{$^2$  Graduate University of Chinese Academy of Sciences,
Yuquan Road 19, Beijing, 100049, China.}
\email{yzfan@pmo.ac.cn (YZF)}

\begin{abstract}
The existence of fast radio bursts (FRBs), a new type of extragalatic transients, has been established recently and quite a few models have been proposed. In this work we discuss the possible connection between the FRB sources and ultra-high energy ($>10^{18}$ eV) cosmic rays. We show that in the blitzar model and the model of merging binary neutron stars, the huge energy release of each FRB central engine together with the rather high rate of FRBs, the accelerated EeV cosmic rays may contribute significantly to the observed ones. In other FRB models including for example the merger of double white dwarfs and the energetic magnetar radio flares, no significant EeV cosmic ray is expected. We also suggest that the mergers of double neutron stars, even if they are irrelevant to FRBs, may play a non-ignorable role in producing EeV cosmic ray protons if supramassive neutron stars were formed in a good fraction of mergers and the merger rate is $\gtrsim 10^{3}~{\rm yr^{-1}~ Gpc^{-3}}$. Such a possibility will be unambiguously tested in the era of gravitational wave astronomy.
\end{abstract}

\keywords{radio continuum: general--- acceleration of particles---cosmic rays}

\setlength{\parindent}{.25in}

\section{Introduction}
In a recent survey for pulsars and fast transients, \citet{Thornton2013} have confirmed \citet{Lorimer2007} and \citet{Keane2012}'s discovery by uncovering four millisecond-duration radio bursts (hereafter FRB) all more than $40\,^{\circ}$ from the Galactic plane. Current data
favor celestial rather than terrestrial origin, and the host galaxy and
intergalactic medium models suggest that they have cosmological redshifts ($z$) of 0.5 to 1 and distances
of up to $\sim 3$ Gpc. The millisecond-duration suggests that central engine is likely either a neutron star or a stellar-mass black hole.
Currently quite a few models have been proposed to interpret FRBs, including the mergers of binary neutron
stars \citep{Hansen2001,Totani2013,Lipunov2014}, energetic magnetar radio flares \citep{Popov2007}, delayed collapses of supramassive neutron stars (SMNS) to black holes (i.e., the so-called blitzar model \citep{Falcke2013}, a highly relevant hypothesis is the possible connection between Gamma-ray Bursts and FRBs \citep{Bannister2012,Zhang2014,Ravi2014}),  mergers of binary white dwarfs \citep{Kashiyama2013}, and flaring stars if FRBs are instead in the Galaxy \citep{Loeb2013}. The discussion of the advantages and disadvantages of these models can be found in \citet{Kulkarni2014} and is beyond the scope of this work. The rate of FRBs is high up to $\sim 10^{4}~{\rm sky^{-1}~day^{-1}}$ \citep{Thornton2013}. If huge amount of energy was released into the circum interstellar medium by the central engine of FRBs, ultra-high energy cosmic rays may be accelerated. The estimate of possible contribution of FRB central engines in producing ultra-high ($>10^{18}$ eV) energy cosmic rays is the main purpose of this work.

\section{Possible high-energy cosmic ray acceleration: model-dependent estimate}
In this section we focus on the cosmological models and in particular the blitzar model, the merging double neutron star model, and model of merger of binary white dwarfs. This is because the total energy released in the models of magnetar giant radio flares\footnote{The newly-born magnetars with initial spin rates close to the centrifugal breakup limit had been suggested to be possible $\sim 10^{20}$ eV cosmic ray sources \citep{Arons2003}. At present it is unclear whether the magnetars generating the giant fast radio bursts initially rotated so quick or not.} \citep{Popov2007} and Galactic flaring stars \citep{Loeb2013} are too small to be sufficient to accelerate and then account for a non-ignorable fraction of ultra-high energy cosmic rays.

\subsection{The blitzar model}\label{sec:II-A}
In the blitzar model, the FRBs were triggered by the collapse of the SMNSs to black holes. As a result of the collapse of SMNS, the magnetic-field lines will snap violently. Accelerated electrons from the traveling magnetic shock dissipate a
significant fraction of the magnetosphere and produce a massive radio burst that is observable out to cosmological distances \citep{Falcke2013}.
The difference between SMNS and normal NS is that the former has to rotate very quickly to not collapse since the gravitational mass of SMNS is larger than that allowed for a non-rotating NS \citep{Friedman1986}. The rapid uniform rotation can enhance the maximum gravitational mass by a factor of $\sim 0.05(P_0/1~{\rm ms})^{-2}$ and thus help make the SMNS stable \citep{Friedman1986}. In addition to the core-collapse supernovae (ccSNe) proposed in \citet{Falcke2013}, the merger of some binary neutron stars can also produce SMNS or even normal NS if the equation of state of the NS material is stiff enough \citep[e.g.,][]{Davis1994,Dai1998a,Dai1998b,Shibata2000,Baumgarte2000,Gao2006,Metzger2008,Zhang2013}. Usually the nascent NSs formed in both ccSNe and merger of double NSs are differentially rotating and the differential rotation is suggested to be more efficient to keep the SN stable. However, the differential rotation is expected to be terminated by the magnetorotational instability as well as magnetic braking \citep{Cook1992,Cook1994,Hotokezaka2013} very quickly. That is why we only consider the effect of uniform rotation in stabilizing the SMNS. Here we concentrate on ccSN-formed SMNS and will discuss the merger-formed SMNS in some detail in section \ref{sec:II-B}.

Before discussing the high-energy cosmic ray acceleration, we examine whether the blitzar model can account for the observed dispersion measures or not. The SN outflow likely has a total mass $M_{\rm sh}\sim~{\rm a~ few}~M_\odot$, and the observed dispersion measures ${\rm DM}\sim 10^{21}~{\rm cm^{-2}}$ \citep{Thornton2013,Lorimer2007} require that the SN outflow should reach a radius $R_{\rm sh}> (M_{\rm sh}/4\pi m_{\rm p} (1+z){\rm DM})^{1/2}\sim 10^{18}~{\rm cm}~(M_{\rm sh}/10~M_\odot)^{1/2}(1+z)^{-1/2}$, where the term $(1+z)$ is introduced to address the effects of both the cosmological time dilation and the frequency shift on the measured DM. Clearly, at such a huge radius the SN shell is also transparent for the 1.3 GHz FRB. The velocity of the SN shell can be estimated as $V_{\rm sh} \sim (E_{\rm inj}/M_{\rm sh})^{1/2}\sim 10^{9}~{\rm cm~s^{-1}}~(M_{\rm sh}/10~M_\odot)^{-1/2}~(E_{\rm inj}/10^{52}~{\rm erg})^{1/2}$. The life of the SMNS should satisfy $t_{\rm life}>R_{\rm sh}/V_{\rm sh}\sim 10^{9} s~(1+z)^{-1/2}(M_{\rm sh}/10~M_\odot)~(E_{\rm inj}/10^{52}~{\rm erg})^{-1/2}$, suggesting a dipole magnetic field strength $B_\perp<1.3\times 10^{12}~{\rm G}~(1+z)^{1/4}(I/1.5\times 10^{45}~{\rm g~cm^{2}})^{3/4}(R_{\rm s}/10^{6}~{\rm cm})^{-3}(P_0/1~{\rm ms})^{1/2}(M_{\rm sh}/10~M_\odot)^{-1/2}$, where $B_{\rm \perp}=B_{\rm s} \sin \alpha$, $B_{\rm s}$ is the surface magnetic field strength at the pole, $R_{\rm s}$ is the radius of the NS and $\alpha$ is the angle between the rotational and dipole axes. Hence the SMNS should not be significantly magnetized otherwise the FRB can not be accounted for (see also \citet{Falcke2013}). We denote such a request as Request-I. Another highly-related request is that the gravitational wave radiation should be weak enough to not dominate over the dipole radiation of the pulsar (i.e., Request-II). In terms of the ellipticity ($\epsilon$) of the pulsar, we need $\epsilon<5\times 10^{-6}(I/1.5\times 10^{45}~{\rm g~cm^{2}})^{-1/2}(P_0/1~{\rm ms})^{2}(t_{\rm life}/10^{9}~{\rm s})^{1/2}$ \citep{Sharpiro1984}. Alternatively, for a rapidly-rotating pulsar, some instabilities, for example the Chandrasekhar-Friedman-Schutz instability \citep{Chandrasekhar1970,Friedman1978}, may occur when the ratio ${\cal R}={\cal T}/|W|$ of the rotational
kinetic energy ${\cal T}$ to the gravitational binding energy $|W|$ is sufficiently large. In the Newtonian
limit, the $l = m = 2$ $f-$mode, which has the shortest
growth time of all polar fluid modes (i.e., $\tau_{\rm GW}\sim 5\times 10^{-6}~{\rm s}~(R_{\rm s}/10^{6}~{\rm cm})^{4}({\cal R}-{\cal R}_{\rm sec})^{-5}$, see Lai \& Sharpiro 1995), becomes
unstable when ${\cal R}\gtrsim {\cal R}_{\rm sec}\approx 0.135$.
%and then carry away the rotational energy via the gravitational radiation (see e.g., Lai \& Sharpiro 1995; Corsi \& M\'{e}sz\'{a}ros 2009 for extended discussion).
Hence ${\cal R}-{\cal R}_{\rm sec}\lesssim 10^{-3}$ is needed to satisfy $\tau_{\rm GW}>t_{\rm life}\sim 10^{9}~{\rm s}$ otherwise the SMNSs collapsed too early to get an sufficient small ${\rm DM}$. Finally, the contribution to the ${\rm DM}$ by the circum-burst medium is less clear. In the estimate of the wind bubble structure at the end of the Wolf-Rayet stage of the massive star, Chevalier et al. (2004) assumed that the surrounding medium has a density typical of the hot, low-density phase of a starburst galaxy (i.e., $n\sim 0.2~{\rm cm^{-3}}$). If it is the case, the contribution to the ${\rm DM}$ by the surrounding medium can be ignored. However, the ccSN may be born in molecular cloud with a typical number density $n_{\rm cloud}\sim 10^{2}~{\rm cm^{-3}}$ and a size of $R_{\rm cloud}\sim (3M_{\rm cloud}/4\pi n_{\rm cloud}m_{\rm p}c^{2})^{1/3}\sim 10~{\rm pc}~(M_{\rm cloud}/10^{4}~{\rm M_\odot})^{1/3}(n_{\rm cloud}/10^{2}~{\rm cm^{-3}})^{-1/3}$, which can give rise to an observed ${\rm DM}\approx 10^{3}~(1+z)^{-1}(M_{\rm cloud}/10^{4}~{\rm M_\odot})^{1/3}(n_{\rm cloud}/10^{2}~{\rm cm^{-3}})^{2/3}~{\rm pc~cm^{-3}}$. Therefore the surrounding molecular cloud should not be very massive/dense (i.e., $M_{\rm cloud}< 10^{4}~{\rm M_\odot}$ for $n_{\rm cloud}\sim 10^{2}~{\rm cm^{-3}}$) otherwise the central ccSNe could not be viable progenitors of the observed FRBs. Such a request is denoted as Request-III. All these three requests impose some challenges on the ccSN scenario in the blitzar model. Nevertheless, the blitzar model can explain some aspects of FRBs \citep[e.g.,][]{Falcke2013} and is still widely adopted in the literature. Below we discuss the possible high energy cosmic ray acceleration in such a scenario.

In view of the sensitive dependence of stabilization on $P_{\rm 0}$, SMNS is unlikely to exist for $P_0>2$ ms unless there is the fine tuning that the mass of SMNS is just tiny above that allowed by the non-rotating NS. The rotational kinetic energy of SMNSs is quite large, i.e., $E_{\rm SN,r}\approx 3\times 10^{52}~{\rm erg}~(I/1.5\times10^{45}~{\rm g~cm^{2}})(P_0/1~{\rm ms})^{-2}$, where $I$ is the moment of inertia.
Before collapsing into a black hole, the SMNS should have lost its rotational energy mainly via magnetic dipole radiation and possibly also gravitational wave radiation \citep{Usov1992,Duncan1992,Fan2013ApJL}. The amount of energy injected into the surrounding medium is $E_{\rm inj} \sim E_{\rm SN,r}/2 \sim 1.5\times 10^{52}~{\rm erg}~(I/1.5\times10^{45}~{\rm g~cm^{2}})(P_0/1~{\rm ms})^{-2}$, if the gravitational wave radiation is not dominant. The frequency of the electromagnetic wave is $\sim 10^{3}$ Hz, which is much lower than the surrounding plasma's frequency $\omega_{\rm p}\sim 5.6\times 10^{4}~{\rm Hz}~n_{\rm e}^{1/2}$, where $n_{\rm e}$ is the number density of the free electrons in the plasma (i.e., the SN outflow) and can be estimated as $\sim 10^{4}~{\rm cm^{-3}}~(M_{\rm sh}/10~M_\odot)({\cal R}_{\rm sh}/0.1)^{-1}(R_{\rm sh}/10^{18}~{\rm cm})^{-3}$, where the width of the outflowing material shell is taken to be a fraction ${\cal R}_{\rm sh}\sim 0.1$ of the $R_{\rm sh}$. Therefore the  electromagnetic wave will be ``trapped" by the plasma. The pressure of the electromagnetic wave outflow is so high that can work on the surrounding
plasma and then the magnetic wind energy will be transferred into the kinetic energy of  the outflow, as indicated by the shallow decay of some GRB afterglows \citep{Dai1998a,Dai1998PRL,Zhang2006}
and the light curves of some superluminous supernovae \citep{Kasen2010,Woosley2010,Inserra2013,Nicholl2013}.
Energetic forward shocks will be driven, and then accelerate protons and other charged particles to ultra-high energies
\footnote{The above estimate is for the assumption that at a radius $R_{\rm sh}\geq 10^{18}~{\rm cm}$ the SMNS wind is still Poynting-flux dominated. If instead the SMNS wind at such a large distance is electron/positron pair dominated and the Poynting-flux is a tiny amount (for example $<0.1$) of the total \cite[e.g.,][]{Kennel1984}, most of the wind energy will be converted into radiation and the acceleration of the high energy cosmic rays by the supernova outflow is less significant.\label{footnote-2}}. Hence in the blitzar model FRBs may be promising sources of EeV cosmic ray protons.

So far, no reliable $\gamma-$ray/X-ray/optical counterparts of FRBs has been identified and the nature of the central engine can not be pinned down. A possible test of the blitzar model is to observe the very early radio emission of GRBs since at least for some GRBs the central engine may be SMNSs (see \cite{Zhang2014} and the references therein). Intriguingly, tentative association of two single dispersed millisecond radio pulses with two GRBs has been reported and these two single dispersed millisecond radio pulses were detected in the few minutes following two GRBs and the arrival times of both pulses are found to coincide with breaks in the GRB X-ray light curves, which likely label the phase transition of the central engine, i.e., the collapse of the SMNS to stellar black hole \citep{Bannister2012}. Such a correlation between FRBs and GRBs, if confirmed in the future, will be in support of the blitzar model.\footnote{As already mentioned in above paragraph, to be a valid source of the observed FRBs, the SMNSe formed in the normal ccSNe should have a typical dipole magnetic field strength $B_\perp\lesssim 10^{12}$ Gauss while the SMNS candidates found in both long and short GRBs usually have $B_\perp> 10^{14}$ Gauss. The energy available for generating FRBs of the GRB SMNSs is about 4 orders of magnitude larger than that of normal ccSN SMNSs.
%So the similarity between the GRB associated FRBs and the normal ccSN associated FRBs is not straightforwardly guaranteed.
Some differences between the FRBs from these two different groups of central engines may be expected. On the other hand, the dipole radiation timescale of such FRB pulsars are so long that at early times the associated supernovae are still ``normal" with a typical energy $\ll E_{\rm inj}\sim~{\rm a~few}\times 10^{52}$ erg, {\it which do not belong to the so-called hypernovae}.}

Now we discuss the cosmic ray particle generation by the SMNS wind-accelerated outflow of the ccSNe. The progenitor star of a ccSN would experience significant mass loss stage and then the surrounding medium is usually not a simple free-wind structure or a constant density structure. For simplicity, here we adopt the structure shown in Fig.1 of \cite{Chevalier2004} to estimate the maximum energy of the protons accelerated at the shock front of the SMNS wind-driven outflow. In such a scenario, significant particle acceleration takes place at $R\sim 5\times 10^{18}$ cm, where is the supergiant shell with a number density $n\sim 10^{2}-10^{3}~{\rm cm^{-3}}$. Following \cite{Bell2001}, the maximum energy of the protons accelerated by the forward shock can be estimated as
\begin{equation}
\varepsilon_{\rm p,M}^{\rm ccSN}\sim 10^{18}~ {\rm eV}~({V\over 10^{9}~{\rm cm~s^{-1}}})^{2}({n\over 10^{2}~{\rm cm^{-3}}})^{1/2}({\epsilon_{\rm B}\over 10^{-2}})^{1/2},
\label{eq:E_max1}
\end{equation}
where $\epsilon_{\rm B}$ is the fraction of shock energy given to magnetic field and the velocity of the SN shell has been estimated to be $V_{\rm sh} \sim (E_{\rm inj}/M_{\rm sh})^{1/2}\sim 10^{9}~{\rm cm~s^{-1}}~(M_{\rm sh}/10~M_\odot)^{-1/2}~(E_{\rm inj}/10^{52}~{\rm erg})^{1/2}$.
The charged particles reach energies larger by a factor of $Z$, the charge number. The magnetic field generated by the shock is $B\sim 10^{-3}~{\rm Gauss}~(\epsilon_{\rm B}/10^{-2})^{1/2}(n/10^{2}~{\rm cm^{-3}})^{1/2}(V_{\rm sh}/10^{9}~{\rm cm~s^{-1}})$, which is too low to effectively cool the accelerating EeV cosmic rays.

The rate of FRBs is ${\cal R}_{_{\rm FRB}} \sim 10^{-3}~{\rm year^{-1}}$ per galaxy \citep{Thornton2013}, which is about one order of magnitude lower than the ccSN rate ${\cal R}_{\rm ccSN}\sim 10^{-2}~{\rm year^{-1}}$ per galaxy.
If FRBs are indeed the cry of the dying SMNSs, the energy released into the surrounding material of each SMNS is $E_{\rm inj}\sim 10^{52}$ erg, implying that the total energy input by FRB sources is comparable to the input by all other ccSNs, and thus FRBs should be one kind of the main sources of cosmic rays.
In particular, as found in eq.(\ref{eq:E_max1}), the most energetic cosmic ray protons can reach the energy of $10^{18}$ eV, and might be the dominant component at such energies.
The injected cosmic-ray density
\begin{equation}
\dot{\epsilon}_{\rm CR}\gtrsim 10^{47}\eta\omega({E_{\rm inj}\over 10^{52}~{\rm erg}})({{\cal R}_{\rm FRB}\over 10^{4}~{\rm yr^{-1}~ Gpc^{-3}}})~{\rm erg}~ {\rm yr^{-1}}~{\rm Mpc^{-3}}
\end{equation}
per energy decade
%for each energy input $\sim E_{\rm inj}\sim 10^{52}$ erg
with $\eta$ as the cosmic ray acceleration efficiency, and $\omega=1/\ln(\varepsilon_{\rm max}/\varepsilon_{\rm min})\simeq 0.1$ as the fraction of the total cosmic ray energy at each energy decade.
Then the corresponding observed cosmic ray flux at $\sim 10^{18}$ eV is
\begin{eqnarray}
F_{\rm EeV-CR} &\sim & 10^{-28} \eta({\omega\over0.1})({E_{\rm inj}\over 10^{52}~{\rm erg}})\nonumber\\
&&({R_{\rm FRB}\over10^{4}~{\rm yr^{-1}~ Gpc^{-3}}})~{\rm m^{-2}~s^{-1}~sr^{-1}~eV^{-1}}.
\label{eq:F-EeV-CR}
\end{eqnarray}
%where $f_z$ is the correction factor for the contribution from high-redshift sources.
Roughly, a rate of ${\cal R}_{\rm FRB}\sim 10^{4}~{\rm yr^{-1}~ Gpc^{-3}}$ would result in the observed flux of cosmic-ray as
$10^{-28}\eta ~{\rm m^{-2}~s^{-1}~sr^{-1}~eV^{-1}}$,
which can meet the observed cosmic ray flux $F_{\rm obs}(\varepsilon)=C(\varepsilon/6.3\times10^{18}~\rm eV)^{-3.2\pm0.05}$ with $C=(9.23\pm0.065)\times10^{-33}\rm m^{-2}~s^{-1}~sr^{-1}~eV^{-1}$ \citep{Nagano2000},
for the EeV cosmic ray acceleration efficiency $\eta \sim 0.03$.

EeV cosmic rays in principle can produce PeV neutrinos via interacting with the interstellar medium, but we do not expect significant PeV neutrino emission since the $>10^{17}$ eV protons can not be effectively confined and the energy loss via pion production is ignorable unless the FRBs were born in the starburst galaxies and in particular the so-called ultra-luminous infrared galaxies \citep{He2013}.

\subsection{The model of merging double Neutron Stars}\label{sec:II-B}
In the model of merging double neutron stars for FRBs, the radiation mechanism may be coherent radio emission, like radio pulsars, by magnetic braking when magnetic fields of neutron stars are synchronized to binary rotation at the time of coalescence \citep{Totani2013}.
In addition to FRBs, the mergers of binary neutron stars may give rise to short GRBs or other kinds of violent explosions
with possible central engines of magnetized millisecond neutron stars \citep[e.g.,][]{Duncan1992, Davis1994,Dai1998a,Baumgarte2000,Shibata2000,Duez2006, Price2006, Rosswog2007,Giacomazzo2013}.
The latest numerical simulations suggest that SMNS can be formed in the merger of a NS binary with $M_{\rm tot}\sim 2.6~M_\odot$ (note that among the ten NS binaries identified so far, five systems have such a total gravitational mass \citep{Lattimer2012}) for reasonably stiff equation of states that are favored by current rest mass measurements of pulsars (see \citet{Hotokezaka2013} and \citet{Fan2013PRD} and the references therein). There are growing, though inconclusive, observational evidences for forming SMNS or even stable NS in NS-NS mergers. The most-widely discussed one in the literature is the X-ray plateau followed by an abrupt
cease in the afterglow light-curve of many short GRBs \citep{Gao2006,Metzger2008,Rowlinson2010,Rowlinson2013}.
In addition, the observations on short bursts such as GRB 051221A and GRB 130603B indicate an energy injection from a highly-magnetized SMNS and the injected energy is as high as $10^{51}-10^{52}{\rm erg}$ \citep{FanXu2006,Rowlinson2010,DallOsso2011,Fan2013ApJL}.
Moreover, as pointed out firstly by \citet{FanXu2006}, usually the material ejected during the binary neutron star merger is not expected to be more than $\sim 0.01~M_\odot$ and could be accelerated to a mildly-relativistic velocity by the wind of SMNS and then produce  X-ray/optical/radio afterglow emission (see \cite{Zhang2013b} for detailed numerical calculation of the lightcurves), which can well account for the cosmological relativistic fading source PTF11agg, a remarkable event not associated with a high energy counterpart \citep{Wang2013,Wu2013}.
It thus seems reasonable to assume that SMNSs, which likely collapsed at $t_{\rm c}\sim 10^{2}-10^{4}$ s \citep{Rowlinson2013}, were formed in a good fraction of NS-NS mergers.

\begin{figure}
\includegraphics[width=80mm,angle=0]{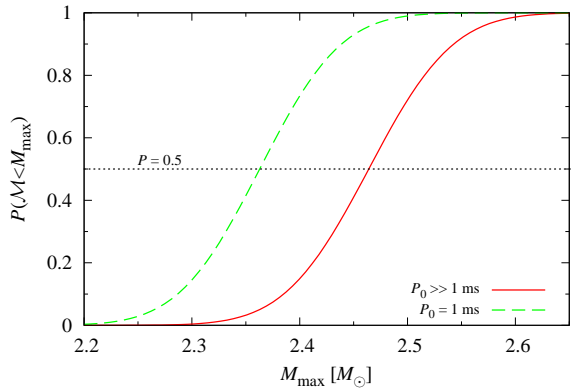}
\caption{The possibility distribution, as a function of $M_{\rm max}$, of forming supramassive ($P_0=1$ ms) neutron stars or stable neutron stars ($P_0\gg 1$ ms) in the mergers of binary neutron stars.} \label{fig:NSs}
\end{figure}

{The prospect of forming SMNS in the mergers can also be roughly estimated as the following. On the one hand, the gravitational mass
($M$) of the isolated neutron star is related to the baryonic mass ($M_{\rm b}$) as
$M_{\rm b}\approx M+\alpha M^{2}$, where $\alpha\approx 0.08~M_\odot^{-1}$ \citep{Lattimer1989,Timmes1996}. On the other hand, in the numerical simulations of mergers of binary neutron stars performed in full general relativity
incorporating the finite-temperature effect and neutrino cooling, \citet{Sekiguchi2011} found that the effect of the thermal energy is significant and can increase the maximal gravitational mass $M_{\rm max}$ by a factor of $20\%-30\%$ for a high-temperature state with $T\geq 20$ MeV. Since they are not supported by differential rotation, the supermassive remnants
were predicted to be stable until neutrino cooling, with luminosity of $\gtrsim 3\times 10^{53}~{\rm erg~s^{-1}}$, has
removed the pressure support in a few seconds \citep{Sekiguchi2011}. After the neutrino cooling, the supermassive remnant is still stable if
\[\alpha
M_{\rm r,max}^{2}+M_{\rm r,max}-(M_{1}+M_{2})-\alpha (M_{1}^{2}+M_{2}^{2})+m_{\rm loss}>0,
\]
where $m_{\rm loss}$ is the baryonic mass loss of the system during the merger, $M_{\rm r,max}\approx [1+0.05(P_0/1~{\rm ms})^{-2}]M_{\rm max}$, and $M_1$ and $M_2$ are the gravitational masses of the binary neutron stars, respectively. $m_{\rm loss}\sim 10^{-3}-10^{-2}~M_\odot$ has been inferred in the numerical simulation \citep{Rosswog1999}. In the modeling of the infrared bump of short GRB 130603B, $m_{\rm loss}\sim~{\rm a~few}\times 10^{-2}~M_\odot$ is needed \citep{Tanvir2013,Berger2013,Fan2013ApJL}. As a conservative estimate we take $m_{\rm loss}=0.01~M_\odot$.
In order to estimate the mass distribution of the neutron stars in the NS-NS binary systems, \citet{Ozel2012} divided the sample into one of pulsars and
one of the companions (For the double pulsar system J0737-3039A, they
assigned the faster pulsar to the ``pulsar" and the slower to
the ``companion" categories). Repeating the above inference
for these two subgroups individually, \citet{Ozel2012} found that the most
likely parameters of the mass distribution for the pulsars are
$M_0 = 1.35M_\odot$ and $\sigma = 0.05~M_\odot$, whereas for the companions
they are $M_0 = 1.32M_\odot$ and $\sigma= 0.05M_\odot$. Hence in our simulation, the distributions of gravitational masses of neutron stars as $dN_{\rm NS}/dM \propto \exp[-(M-M_0)^2/2\sigma^2]$ with these parameters are adopted. The possibility distribution of the gravitational masses of supermassive remnants (i.e., ${\cal M}\approx {[-1+\sqrt{1+4\alpha[M_1+M_2+\alpha (M_1^{2}+M_2^{2})-m_{\rm loss}}
]/2\alpha}$) formed in the simulated double neutron star mergers is presented in Fig.\ref{fig:NSs}. We find for $M_{\rm max}\geq 2.36~M_\odot$, about half of the mergers will produce SMNSs with  $P_{0}=1~{\rm ms}$. Observationally the pulsar PSR J0348+0432 has an accurately
measured gravitational mass $2.01\pm0.04~M_\odot$ \citep{Antoniadis2013} and J1748-2021B has a gravitational mass $\approx 2.74\pm 0.21~M_\odot$ \citep{Lattimer2012}. Hence $M_{\rm max}\sim 2.36~M_\odot$ is still possible, with which a sizeable fraction of NS-NS mergers may produce SMNSs.
}

%*****************************Fig.1***************************************
%*************************************************************************

We have briefly mentioned in section \ref{sec:II-A} that {\it in the blitzar model} a small fraction of FRBs may be relevant to the mergers of double neutron stars that produce SMNS remnants. In such a scenario some FRBs are expected to be detected in the afterglow emission phase of short GRBs. {\it In the model of merging double neutron stars},
FRBs are generated at the time of coalescence of double NSs and are expected to {\it precede} short GRBs or other kinds of violent explosions (i.e., no FRB is expected to occur in the afterglow phase of short GRBs) and the three requests outlined in Sect. \ref{sec:II-A} do not apply. The follow-up observations of FRBs and short GRBs would be crucial to distinguish between the blitzar model and the merging neutron star model. The origin of merging double neutron stars for FRBs, if confirmed in the future, has interesting implication on the sources of EeV cosmic ray protons. This is because there are some tentative evidence for the formation of SMNSs in a sizeable fraction of NS-NS mergers and such a kind of long-lived central engine can accelerate the materials ejected during the merger to very high velocities \citep{FanXu2006,Zhang2013b,Wu2013}. {The physical reason for converting the SMNS wind energy into the kinetic energy of the forward shock is the same as that in the case of ccSNe. Within a radius $R_{\rm sh}\lesssim ct_{\rm c}\sim 3\times 10^{14}~{\rm cm}~(t_{\rm c}/10^{4}~{\rm s})$, the SMNS wind is likely Poynting-flux dominated rather than electron/positron pair dominated \citep{Vlahakis2004}. The frequency of the electromagnetic wave of the SMNS ($\sim 10^{3}$ Hz) is much smaller than the surrounding plasma's frequency $\omega_{\rm p}\sim 5.6\times 10^{4}~{\rm Hz}~n_{\rm e}^{1/2}$, where $n_{\rm e}\sim 3\times 10^{11}~{\rm cm^{-3}}~(M_{\rm ej}/0.01~M_\odot)({\cal R}_{\rm sh}/0.1)^{-1}(R_{\rm sh}/3\times 10^{14}~{\rm cm})^{-3}$ is the number density of the free electrons in the merger outflow with a rest mass $M_{\rm ej}\sim 0.01~M_\odot$. As a result, the electromagnetic wave is ``trapped" by the plasma. The high magnetic pressure works on and hence accelerate the surrounding plasma (see Yu et al. 2013 and the references therein).} The SMNS-driven outflow, almost isotropic, will generate energetic forward shocks and then accelerate ultra-high energy cosmic rays.

The maximum energy of the cosmic rays accelerated by the wide outflow driven by the SMNS can be estimated as
\begin{eqnarray}
\varepsilon_{\rm p,M}^{\rm NS-NS} &\sim & \beta Z e B_{\rm d} R_{\rm d} \nonumber\\
&\sim & 1.5\times 10^{18}~ {\rm eV}~Z({\beta \over 0.5})^2({E_{\rm inj}\over 10^{52}~{\rm erg}})^{1\over 3}\nonumber\\
&&({n\over 10^{-2}~{\rm cm^{-3}}})^{1\over 6}({\epsilon_{\rm B}\over 10^{-2}})^{1\over 2}{\Gamma}^{1\over 3},
\label{eq:E_max2}
\end{eqnarray}
where $\beta$ is the velocity of the ejecta in units of the speed of light $c$, $e$ is the electron's charge, $R_{\rm d}\sim 1.2\times 10^{18}~{\rm cm}
~({E_{\rm inj}/10^{52}~{\rm erg}})
^{1/3}(n/10^{-2}~{\rm cm^{-3}})^{-1/3}(\Gamma/5)^{-2/3}$ is the deceleration radius, and $B_{\rm d}=0.02~{\rm Gauss}~ \beta (\Gamma/5) (n/0.01~{\rm cm^{-3}})^{1/2}(\epsilon_{\rm B}/0.01)^{1/2}$ is the magnetic field strength at $R_{\rm d}$. The $\beta$ has been normalized to $0.5$ since in the presence of a highly magnetized SMNS, the material ejected during the NS-NS merger is expected to be accelerated to a trans-relativistic or even mildly-relativistic velocity \citep{FanXu2006}. Different from the ccSN scenario, we normalize $n$ to the value of $10^{-2}~{\rm cm^{-3}}$ to address the fact that some mergers of NSs are expected to take place in low density medium. Again the cooling of the accelerating EeV cosmic rays do not suffer significant energy loss via synchrotron radiation due to the low $B_{\rm d}$.

How large is $E_{\rm inj}$? The answer is somewhat uncertain since the SMNS formed in double neutron star mergers might suffer significant energy loss in addition to the regular dipole magnetic radiation. For GRB 130603B displaying a SMNS signature, $E_{\rm inj}\gtrsim 2\times 10^{51}$ erg is needed to account for the multi-wavelength afterglow data. In a good fraction of short GRBs with distinguished X-ray afterglow plateau, as reported in \citet{Rowlinson2013}, the energy release by the central SMNS is found to be $E_{\rm inj}<10^{52}$ erg. The inferred $E_{\rm inj}\sim {\rm a~few}\times 10^{51}$ erg is also favored by the weak radio afterglow emission of most short GRBs and has been suggested to be the signature of the significant gravitational wave radiation of the SMNS \citep{Fan2013ApJL,Fan2013PRD}, {which may be possible if the interior toroidal magnetic field was high up to $\sim 10^{17}$ Gauss that can give rise to a sizeable deformation $\epsilon \sim 0.01$ of the magnetar or the secure instability occurred with a ${\cal R}\sim {\cal R}_{\rm sec}+0.03 \sim 0.165$}. While in the modeling of GRB 051221A and PTF11agg, $E_{\rm inj}\sim 10^{52}$ erg is needed. On average $E_{\rm inj}$ is likely $\sim{\rm quite~a~few}\times 10^{51}$ erg. With  Eq.(\ref{eq:F-EeV-CR}), it is straightforward to show that $\eta \sim 0.1~(E_{\rm inj}/3\times 10^{51}~{\rm erg})^{-1}$ is needed otherwise the accelerated ultra-high energy cosmic rays can not account for a sizeable fraction of the observed EeV ones.
We then suggest that in the model of merging double NSs \citep{Totani2013}, the FRBs may still have significant connection with ultra-high energy cosmic ray sources though the argument is less direct than in the blitzar model \footnote{The additional assumptions are either SMNSs are formed in a considerable fraction of binary neutron star mergers or alternatively the ejection of material with velocities $>0.3c$ during the merger is very important, as hinted by current short GRB observations and by latest numerical simulation.}. Possible byproducts are high energy neutrinos if there are dense/energetic seed photons. Even with very optimistic assumptions, the resulting PeV neutrinos are likely too weak to give rise to significant detection for IceCube like detectors
(see \cite{Gao2013PRD} for a relevant estimate).

\subsection{The model of merger of binary white dwarfs}\label{sec:II-C}
In the model of merger of binary degenerate white dwarfs, the FRBs were produced by coherent emission from the polar region of a rapidly rotating, magnetized massive white dwarf formed after the merger \citep{Kashiyama2013}. The energy budget of the nascent massive white dwarf can be estimated as $E_{\rm bug} \sim G M_{_{\rm WD}}^2/R_{_{\rm WD}} \sim 3\times 10^{50}~{\rm erg} ~(M_{_{\rm WD}}/1~M_\odot)^{2}(R_{_{\rm WD}}/10^{9}~{\rm cm})^{-1}$. Magnetic activity of the post-merger object has been demonstrated
by recent numerical simulations \citep{Ji2013}, in which the magnetic energy of the remnant at its
peak is found to exceed $10^{48}$ erg (the corresponding volume-averaged magnetic field strength
is $\bar{B}\sim 10^{11}~{\rm Gauss}$) and about $M_{\rm ej,_{\rm WD}} \sim 10^{-3}~M_\odot$ mass is ejected
from the system over the run time of the simulations, i.e., $t = 2 \times 10^{4}$ s. With a spin-down luminosity of the magnetized
massive white dwarf $L_{\rm dip,_{\rm WD}} \sim 2\times 10^{38} (B_{\perp,_{\rm WD}}/10^{9}~{\rm G})^{2}$, the ejected material as well as the swept circum medium with a density $n\sim 0.01~{\rm cm^{-3}}$ can not be accelerated to a velocity larger than $\sim 10^{9}~{\rm cm~s^{-1}}$, where we have normalize $B_{\perp,_{\rm WD}}$ to a value of $10^{9}$ G, the surface magnetic field strength of the highly magnetized white dwarfs observed so far. With eq.(\ref{eq:E_max2}) we find that the cosmic ray protons more energetic than $\sim 10^{16}$ eV can not be accelerated, which are not of our interest. If some FRBs are associated with type Ia supernovae, as suggested in \citet{Kashiyama2013}, the supernova outflow can accelerate cosmic rays too. However, it is widely known that SNe Ia outflow can at most accelerate protons to the so-called ``knee" of the cosmic ray spectrum (i.e., $\sim 3~{\rm PeV}$). Therefore in the model of merger of binary white dwarfs, EeV cosmic rays are not expected.

\section{Discussion}
Since the origin of FRBs is still to be pinned down, in this work we have carried out model-dependent estimate on the possible role of FRB sources in accelerating EeV cosmic rays. In the models of magnetar giant radio flares, the merger of binary white dwarfs, and Galactic flaring stars,
significant EeV cosmic ray acceleration is not expected. While in the blitzar model, the cosmological FRB sources are very promising EeV cosmic-ray accelerators thanks to the huge energy release into the surrounding medium by each supramassive neutron star (see Sec. \ref{sec:II-A}). In the model of merging neutron stars, FRBs may still be promising ultra-high energy cosmic ray sources if supramassive neutron stars are formed in a considerable fraction of binary neutron star mergers (see Sec. \ref{sec:II-B}). We also suggest that in the blitzar model the GRB-related FRBs may show some difference from the normal ccSN-related FRBs.
% and the off-beaming GRBs are not good FRB source candidates.

The neutron star mergers, if irrelevant to FRBs, are expected to have a rate lower than ${\cal R}_{_{\rm FRB}}$. Even so, their role in producing EeV cosmic rays may be non-ignorable.
We are aware that the role of NS-NS-merger outflow in accelerating $\leq 10^{17}$ eV cosmic ray protons has been discussed in \citep{Takami2013}, where energy injection and then acceleration of the outflow caused by the SMNS central engine have not been taken into account. However, as summarized in Sec. \ref{sec:II-B}, there are growing evidence for the formation of SMNS in plausibly a non-ignorable fraction of binary neutron star mergers, which help accelerate EeV cosmic rays. For example, inserting the physical parameters inferred from the modeling of GRB 130603B with a SMNS central engine \citep{Fan2013ApJL} into eq.(\ref{eq:E_max2}), we have $\varepsilon_{\rm p,M}^{\rm NS-NS}\sim 2\times 10^{18}$ eV.
Based on extrapolations from observed binary pulsars in the Galaxy, a likely coalescence rate of binary NSs is $\sim 10^{-4}~{\rm yr}^{-1}$ per Milky-Way Equivalent Galaxy \citep{Abadie2010}, which is about one order of magnitude lower than the observed rate of FRBs (The very optimistic estimate of the NS-NS merger rate could be comparable to the rate of FRBs). The beaming-corrected estimates of short GRB rate can be as high as $\sim 10^{3}~{\rm yr^{-1}~ Gpc^{-3}}$ \citep{Coward2012}, and the merger rate of binary NSs is expected to be higher. These two independent estimates are consistent with each other. If each merger injects energy of $\sim 10^{52}$ erg into the surrounding material (i.e., the rotational energy of SMNS is mainly lost via magnetic dipole radiation),
and then drives trans-relativistic shocks ($\beta>0.5$),
the flux of cosmic rays with the energy of $\sim {\rm 10^{18}~eV}$ detectable on the Earth would be
$\sim 10^{-29}\eta (E_{\rm inj}/10^{52}~{\rm erg})~{\rm m^{-2}~s^{-1}~sr^{-1}~eV^{-1}}$,
which can account for $\sim 1/3$ of the observed flux
$\sim 3\times10^{-30} ~{\rm m^{-2}~s^{-1}~sr^{-1}~eV^{-1}}$,
for a reasonable EeV cosmic ray acceleration efficiency $\eta \sim 0.1(E_{\rm inj}/10^{52}~{\rm erg})^{-1}$. As summarized in Sec. \ref{sec:II-B}, if instead $E_{\rm inj}$ is only a few Bethe (i.e., $10^{51}$ erg), as found in GRB 130603B, the binary neutron star mergers may produce only $\sim 10\%$ of the observed EeV cosmic rays. Nevertheless, the importance of NS-NS mergers in producing $>10^{18} ~{\rm eV}$ cosmic rays can be reliably estimated in the foreseeable future, since (i) The gravitational wave observations by advanced LIGO/VIRGO will pin down or impose a tight constraint on the merger rate of binary neutron stars; (ii) The dedicated electromagnetic counterpart searches of the merger events will help us to tightly constrain the total energy injected into the surrounding medium.

In view of the fact that all the Active Galactic Nuclei \citep{Ginzburg1964,Hillas1984}, bright Gamma-ray Bursts \citep{Waxman1995,Vietri1995}  and low-luminosity GRBs \citep{Murase2006,Liu2011}, Type Ic supernovae in particular the so-called hypernovae associated with GRBs  \citep{Dermer2001,Wang2007,Budnik2008,Fan2008,Chakraborti2011,Liu2012}  and clusters of galaxies \citep{Murase2008} can also accelerate cosmic rays to the energies $\geq 10^{18}$ eV, we suggest that the $>10^{18} ~{\rm eV}$ cosmic rays consist of multiple components from different astrophysical sources, possibly including that producing FRBs.\\

\section*{Acknowledgments}
We thank the anonymous referee for helpful comments. This work was supported in part by 973 Programme of China under grants 2013CB837000 and 2014CB845800, National Natural Science of China under grants 11273063, 11303098 and 11361140349, and the Foundation for Distinguished Young Scholars of Jiangsu Province, China (No. BK2012047).  YZF is also supported by the 100 Talents programme of Chinese Academy of Sciences. DMW is partly supported by the Strategic Priority Research Program ``The Emergence of Cosmological Structures" of the Chinese Academy of Sciences, Grant No. XDB09000000. HNH is also supported by China Postdoctoral science foundation under grants 2012M521137 and 2013T60569.\\

\clearpage

\end{document}